\documentclass[preprint,preprintnumbers,amsmath,amssymb,nofootinbib,aps,usenatbib]{revtex4}
\pdfoutput=1
\usepackage{epsfig}
\usepackage{graphicx}
\usepackage{dcolumn}
\usepackage{bm}
\usepackage{amsmath}
\usepackage{amssymb}
\usepackage{latexsym}
\usepackage{color}
\usepackage{hyperref}
\usepackage{mathrsfs}

\newcommand{\be}{\begin{equation}}
\newcommand{\ee}{\end{equation}}
\newcommand{\bea}{\begin{eqnarray}}
\newcommand{\eea}{\end{eqnarray}}

\newcommand{\like}{\mathscr{L}}

 \def\@testdef #1#2#3{%
   \def\reserved@a{#3}\expandafter \ifx \csname #1@#2\endcsname
  \reserved@a  \else
 \typeout{^^Jlabel #2 changed:^^J%
 \meaning\reserved@a^^J%
 \expandafter\meaning\csname #1@#2\endcsname^^J}%
 \@tempswatrue \fi}

\begin{document}


\title{Kinematic reconstruction of torsion as dark energy in Friedmann cosmology}

\author{A. M. Vicente$^{1}$} \email{amvfisico@gmail.com}
\author{J. F. Jesus$^{1,2}$}\email{jf.jesus@unesp.br}
\author{S. H. Pereira$^{1}$} \email{s.pereira@unesp.br}

\affiliation{$^1$Universidade Estadual Paulista (UNESP), Faculdade de Engenharia e Ci\^encias de Guaratinguet\'a, Departamento de F\'isica - Av. Dr. Ariberto Pereira da Cunha 333, 12516-410, Guaratinguet\'a, SP, Brazil
\\
$^2$Universidade Estadual Paulista (UNESP), Instituto de Ci\^encias e Engenharia, Departamento de Ci\^encias e Tecnologia - R. Geraldo Alckmin 519, 18409-010, Itapeva, SP, Brazil
}


\def\zt{\mbox{$z_t$}}

\vspace{1.5cm}
\begin{abstract}
In this paper we study the effects of torsion of space-time in the expansion of the Universe as a candidate to dark energy. The analysis is done by reconstructing the torsion function along cosmic evolution by using observational data of Supernovae type Ia, Hubble parameter {and Baryon Acoustic Oscillation} measurements. We have used a kinematic model for the parameterization of the comoving distance and the Hubble parameter, then the free parameters of the models are constrained by observational data. The reconstruction of the torsion function is obtained directly from the data, using the kinematic parameterizations.

\end{abstract}

\maketitle



\section{Introduction}
\label{sec:introd}

Despite the success of the flat $\Lambda$CDM model, which is consistent with much observational data and correctly predicts the current phase of acceleration of the universe \cite{Farooq:2016zwm, Scolnic:2017caz, Aghanim:2018eyx}, some theoretical and observational discrepancies \cite{Riess:2020sih, Martinelli:2019krf, Bull:2015stt} open the possibility for studying extensions of the standard model or even searching for alternative cosmological and gravitational models. Natural extensions of the standard cosmological model emerge when gravitational theories beyond the Riemannian framework of general relativity are adopted. Einstein-Cartan (EC) gravity is an example that has been recently explored in different contexts \cite{Pasmatsiou:2016bfv, Kranas:2018jdc, Pereira:2019yhu, Barrow:2019bvx, Mehdizadeh:2018smu, Medina:2018rnl, Luz:2019frs, Khakshournia:2019jzs, Marques:2019ifg, Bose:2020mdm, Bolejko:2020nbw, Izaurieta:2020vyh, Cabral:2020mzw, Cruz:2020hkh, Shaposhnikov:2020frq, Shaposhnikov:2020aen, Shaposhnikov:2020gts, Bondarenko:2020kwm, Karananas:2021zkl, Guimaraes:2020drj, Kasem:2020wsp, Pereira:2022cmu, Elizalde:2022vvc, Liu:2023znv, Akhshabi:2023xan,Usman:2023nww,Chakraborty:2023neh,Li:2023ubd}. The main feature of such a theory is that the connection is no longer symmetric, and its antisymmetric part gives rise to the torsion tensor, which naturally enters the field equations and alters the gravitational and cosmological dynamics.

{Torsion effects has been discussed in different cosmological scenery in recent years. The simplest way to take into account the presence of torsion in a homogeneous and isotropic spacetime is through a scalar function $\phi(t)$ that evolves solely with cosmic time. Just to cite some examples, in \cite{Kranas:2018jdc} a specific torsion function was adopted, namely $\phi(t)=\lambda H(t)$, where $H(t)$ is the Hubble parameter. With this ansatz, a strong cosmological bounds on $\lambda$ was obtained  by studing the primordial nucleosynthesis of helium-4, namely  $(-0.005813 < \lambda < +0.019370)$. Following this same approach, \cite{Usman:2023nww} examine the cosmic evolution of the growth
of perturbations with respect to matter content at recent past era and for the early Universe under the framework of non-
zero torsion cosmology. Dark energy effects coming from torsion was studied in \cite{Pereira:2019yhu}, where the torsion function was taking as $\phi(t) = -\alpha H(t)\rho_m(t)^n$, being $\rho_m(t)$ the matter density. It was found that such model is compatible with recent cosmological data for $\alpha = 0.14^{+0.14}_{-0.12}$ and $n=-0.47^{+0.26}_{-0.36}$. The role of matter in Einstein-Cartan gravity was studied in \cite{Karananas:2021zkl}. The contribution of torsion as a dark matter component was studied in two different approaches in \cite{Pereira:2022cmu}, where it was shown that the dark matter sector can be interpreted as an effective coupling of the torsion with ordinary baryonic matter. 
The effects of Einstein-Cartan theory on the propagation of gravitational wave amplitude were studied in \cite{Elizalde:2022vvc}. In \cite{Akhshabi:2023xan}, it was shown that the tension between late-time and early-universe measurements of the Hubble parameter can be alleviated when certain models of torsion is used to determine the Hubble parameter with the observed time delays in gravitational lensing systems.  An inflationary model was proposed in \cite{Guimaraes:2020drj}. Cosmological signatures of torsion were studied in \cite{Bolejko:2020nbw}, and high-energy scattering was calculated in \cite{Bondarenko:2020kwm}.}


Since the explicit form of the $\phi(t)$ function is unknown, as well as its evolution over time, we have two different approaches to address this issue. The first approach involves setting specific forms for $\phi(t)$ and constraining the free parameters of the model. The second approach aims to study its evolution based solely on certain sets of observational data. In other words, what should be the form of $\phi(t)$ in order for it to be compatible with available observational data? {In this study, we aim to reconstruct the torsion function $\phi(t)$ without making assumptions about the specific cosmological dynamics.}

There are two ways to reconstruct the cosmic evolution of a given observable without resorting to a specific dynamic model. The first method is through the so-called cosmographic or kinematic models \cite{kine1, kine2, kine3, kine4, kine5, kine6, kine7}, where a particular parameterization is chosen for the observable, and its free parameters are constrained by observational data. {Cosmological evolution inferences of physical quantities from kinematic observables is the core of cosmographic analysis and its efficiency lies in the ability to test cosmological models that are compatible with the cosmological principle without assumptions about material content of the Universe \cite{Bolotin:2011beg,Zhang:2016urt,rezaei2021,mehrabi2021, velasques2021}. The cosmographic approach with emphasis on the running vacuum models was studied recently in \cite{rezaei2021}. The redshift drift phenomenon was analyzed in \cite{lobo2020}.}

Another approach was recently proposed by Seikel, Clarkson, and Smith \cite{Seikel:2012uu}, where Gaussian Processes based on a non-parametric method are used to reconstruct cosmological observables without the need for a specific parameterization. This method has been employed to reconstruct various observables, including the dark energy equation of state parameter \citep{Holsclaw:2010nb}, luminosity distance \cite{Seikel:2012uu}, transition redshift \cite{Jesus:2019nnk}, the Hubble parameter \cite{Shafieloo:2012ht}, dark energy scalar field potential \cite{Li:2006ea, Sahlen:2005zw}, interaction between dark matter and dark energy \cite{Mukherjee:2021ggf, vonMarttens:2020apn}, the cosmic distance duality relation \cite{Mukherjee:2021kcu}, among others.

In the present work, we employ the Hubble parameter $H(z)$ data, the supernovae type Ia (SNe Ia) data {and the Baryon Acoustic Oscillation (BAO)} to reconstruct the torsion function $\phi(t)$  {and the deceleration parameter $q(z)$, assuming only a kinematic parameterization for the Hubble parameter $H(z)$ and the luminosity distance $D_C(z)$. For constraining the free parameters of the two different parameterizations we have used Markov chain Monte Carlo (MCMC) algorithm. The main advantage of this method is obtaining a direct measure of expansion parameters directly from the data.}

The paper is organized as follows. In Section II the main equations are presented. In Section III dataset is combined with equations of previous section to reconstruct torsion function {and the results are shown. Conclusions are left to Section IV.}

\section{Friedmann cosmology with torsion}
\label{sec:torsion}

{The cosmological equations from EC gravity is constructed by a generalization of the Einstein-Hilbert Action \cite{Ivanov2016,Shapiro:2001rz}:
\begin{equation}
    S = -\frac{1}{\kappa^2}\int d^4x \sqrt{-g}(R-2\Lambda + \mathcal{L}_M)\,,
\end{equation}
with $\kappa^2 = {8\pi G}$, $\Lambda$ the cosmological constant term and $\mathcal{L}_M$ the matter fields lagrangian density. The Ricci scalar $R$ comes from the Ricci tensor  $R_{\mu\nu}$, which is constructed with the generalized connection $\Gamma^\alpha_{~~\mu\nu}=\Tilde{\Gamma}^\alpha_{~~\mu\nu}+K^\alpha_{~~\mu\nu}$, where $K^\alpha_{~~\mu\nu}$ defines the contorsion tensor and $\Tilde{\Gamma}^\alpha_{~~\mu\nu}$ is the usual symmetric Christoffel symbol. The contorsion tensor is related to the torsion tensor $S^{\alpha}_{~~\mu\nu}$ through $K^\alpha_{~~\mu\nu}=S^{\alpha}_{~~\mu\nu} + S_{\mu\nu}^{~~~\alpha} + S_{\nu\mu}^{~~~\alpha}$. Additionally, cosmological principle requires torsion to be related to a homogeneous function $\phi(t)$}, which comes from an ansatz Combining field equations with the cosmological principle, the Friedmann equations in a flat background including a general matter density $\rho$, pressure $p$ and cosmological constant $\Lambda$ are ({see Appendix A for a more complete deduction}):
\begin{align}
        H^2 &=\frac{8\pi G}{3}\rho  + \frac{\Lambda}{3}-4\phi^2 - 4H\phi\,,\label{H2}\\
         \dot{H}+H^2 &=-\frac{4\pi G}{3}(\rho+3p)  + \frac{\Lambda}{3} - 2\dot{\phi} - 2H\phi\,,\label{Hd2}
\end{align}
where $H=\dot{a}/a$ is the Hubble function and $a(t)$ the scale factor of the universe coming from a flat FRW metric, $ds^2 =  -dt^2 + a^2dx_idx^i$. For a barotropic matter satisfying an equation of state of the form $p=w \rho$, the continuity equation reads:
    \begin{equation}
        \dot{\rho}+3(1+w)H\rho + 2(1+3w)\phi \rho = 4\phi\frac{\Lambda}{8\pi G}\,.\label{rhodot}
    \end{equation}

The deceleration parameter, defined by:
\be
q=-\frac{\ddot{a}a}{\dot{a}^2}\,, \label{qa}
\ee
can be written as:
\be
q = \frac{4\pi G}{3}\frac{(1+3w)\rho}{H^2}-\frac{\Lambda}{3H^2} + 2\frac{\dot{\phi}}{H^2} + 2 \frac{\phi}{H}\,.\label{q}
\ee
Given a torsion function $\phi(t)$ the above system of equations can be solved, at least numerically. 

From now on we will assume that torsion assumes the role of the dark energy component, thus we make $\Lambda=0$, with $\rho=\rho_b+\rho_{dm}$ accounting for baryonic and dark matter component, since they are indistinguishable in the background level. {Here we neglect the radiation contribution because we are mainly interested in the late time Universe evolution, when radiation is negligible.} We also change to redshift derivatives in order to obtain the $\phi(z)$ reconstruction. By using $\frac{d}{dt}=-H(1+z)\frac{d}{dz}$ together the assumption of a pressureless matter ($w=0$), Eqs. \eqref{H2} and \eqref{Hd2} can be combined as:
\be
-2H(1+z)\frac{dH}{dz}+3H^2=-4\phi^2+4H(1+z)\frac{d\phi}{dz}-8H\phi
\ee
or
\be
\frac{d\phi}{dz}=\frac{3H}{4(1+z)}-\frac{1}{2}\frac{dH}{dz}+\frac{2\phi}{1+z}+\frac{\phi^2}{H(1+z)}\,.
\label{dphidz}
\ee
Furthermore, we change to dimensionless variables by defining $\Phi\equiv\frac{\phi}{H_0}$ and $E(z)=\frac{H(z)}{H_0}$, where $H_0=H(z=0)$ is the Hubble constant. In this case, \eqref{dphidz} is written as:
\be
\frac{d\Phi}{dz}=\frac{3E(z)}{4(1+z)}-\frac{1}{2}\frac{dE}{dz}+\frac{2\Phi}{1+z}+\frac{\Phi^2}{E(z)(1+z)}
\label{dPhidz}
\ee
For a given $E(z)$ function, \eqref{dPhidz} is a Riccati equation\footnote{See Ref. \cite{Reid1972} for further details.} for $\Phi(z)$, which is a non-linear ordinary differential equation.

In order to solve a Riccati equation, we must to find a particular solution and then use it in order to convert it into a Bernoulli equation, which has a well-known solution. We try to find a particular solution by inspection of Eq. \eqref{dPhidz} by trying $\Phi=-\alpha E$ (or $\phi=-\alpha H$), where $\alpha$ is a constant. From (\ref{dphidz}) we obtain:
\be
\left(\alpha-\frac{1}{2}\right)E'+\left(\alpha^2-2\alpha+\frac{3}{4}\right)\frac{E}{1+z}=0
\ee
where a prime denotes derivative with respect to $z$. The solution for any function $E(z)$ is $\alpha=\frac{1}{2}$, with $\Phi=-\frac{E}{2}$ being a particular solution to the Riccati equation \eqref{dPhidz}. As we have already studied in \cite{Pereira:2019yhu}, the particular solution of the form $\phi=-\alpha H$ for torsion acting as dark energy is not compatible with observational data. In this way, we aim to find another solution of the Riccati equation which is compatible with observational data.

Following the methods of solution of a Riccati equation, if we make a change of variable $\Phi=y-\frac{E}{2}$ into Eq. (\ref{dPhidz}), we obtain the Bernoulli equation:
\be
y'=\frac{y}{1+z}+\frac{y^2}{E(1+z)}\,,
\ee
which, by changing to $v=\frac{1}{y}$, becomes a linear equation:
\be
v'+\frac{v}{1+z}=-\frac{1}{E(1+z)}\,
\ee
which finally has a well-known solution, given by:
\be
v=\frac{c_1}{1+z}-\frac{1}{1+z}\int\frac{dz}{E(z)}\,,
\label{eqv}
\ee
where $c_1$ is an arbitrary constant. Notice that we may write the indefinite integral in \eqref{eqv} in terms of the dimensionless comoving distance {\cite{WangY2009}}:
\be
D_C(z)\equiv\int_0^z\frac{dz}{E(z)}\,
\label{conectaHD}
\ee

It is important to mention that $D_C$ is a dimensionless distance in the sense that $D_C(z)\equiv\frac{d_C(z)}{d_H}$, where $d_C$ is dimensionful comoving distance and $d_H=\frac{c}{H_0}$ is current Hubble distance. {Later soon, in next section, $D_C$ will encode SNe Ia data information \cite{WangY2009}}.

So, Eq. \eqref{eqv} becomes simply:
\be
v=\frac{c_2-D_C(z)}{1+z}\,.
\label{eqvDc}
\ee
{where $c_2$ is another arbitrary constant.} Thus, we have a solution for $\Phi=\frac{1}{v}-\frac{E}{2}$ given by:
\be
\Phi=\frac{1+z}{c_2-D_C(z)}-\frac{E(z)}{2}\,,\label{eq14}
\ee
where the arbitrary constant $c_2 = \frac{2}{2\Phi_0+1}$ can be obtained from the initial condition $\Phi(z=0)=\Phi_0$. Thus, we obtain for (\ref{eq14}):
\be
\Phi=\frac{(2\Phi_0+1)(1+z)}{2-(2\Phi_0+1)D_C(z)}-\frac{E(z)}{2}\,,
\label{Phisol}
\ee
where $\Phi_0$ can be related to the present day values of the density parameters in the following way: We can define a density parameter for torsion from the Friedmann equation \eqref{H2}, by dividing it by $H^2$:
\be
1 = \frac{8\pi G\rho}{3H^2} - \frac{4\phi^2}{H^2} - \frac{4\phi}{H}\,.
\ee

In the same way that we define the matter density parameter as $\Omega_m\equiv\frac{8\pi G\rho}{3H^2}$, we can also define:
\be
\Omega_\phi\equiv - \frac{4\phi^2}{H^2} - \frac{4\phi}{H}=-4\left(1+\frac{\Phi}{E}\right)\frac{\Phi}{E}\,.
\label{Omegaphidef}
\ee
By inserting the solution \eqref{Phisol} into \eqref{Omegaphidef}, we obtain:
\be
\Omega_\phi(z)=1-\frac{4(2\Phi_0+1)^2(1+z)^2}{E(z)^2\left[2-(2\Phi_0+1)D_C(z)\right]^2}\,.
\label{Omegaphiz}
\ee
For a spatially flat universe, $\Omega_\phi=1-\Omega_m$ so:
\be
\Omega_m(z)=\frac{4(2\Phi_0+1)^2(1+z)^2}{E(z)^2\left[2-(2\Phi_0+1)D_C(z)\right]^2}\,,
\label{Omegamz}
\ee
and evaluating \eqref{Omegaphiz} and \eqref{Omegamz} at $z=0$, since $D_C(0)=0$ and $E(0)=1$, we find the relation:
\be
(2\Phi_0+1)^2=1-\Omega_{\phi0}=\Omega_{m0}\,,\label{eq19}
\ee
which shows that the quantity $(2\Phi_0+1)=\pm \sqrt{\Omega_{m0}}$ is directly related to the present day value of the matter density parameter $\Omega_{m0}$. We will choose to work with the negative value of the previous relationship, thus (\ref{Phisol}) can be put in the final form:
\be
\Phi(z)=\frac{-\sqrt{\Omega_{m0}}(1+z)}{2+\sqrt{\Omega_{m0}}D_C(z)}-\frac{E(z)}{2}\,.
\label{Phisolf}
\ee
Notice that the choice of the negative value for $2\Phi_0+1$ avoids a singularity in the denominator of the first term and also maintains the torsion function with a global negative sign, as discussed in \cite{Kranas:2018jdc}. If one knows the current value of the density parameter $\Omega_{m0}$ and knows the evolution of $E(z)$ or $D_C(z)$, one can obtain $\Phi(z)$ from Eq. \eqref{Phisolf}. 

\section{Methodology and Results}

In order to verify the effect of torsion as a candidate to dark energy along cosmic evolution, we have used a parametric regression method to reconstruct the Hubble function $H(z)$ and the comoving distance $D_C(z)$. {Their free parameters were then constrained by using two different sets} of observational data. The first are the {Pantheon+\&SH0ES sample \cite{pantheon+}, consisting of 1701 light curves for 1550 distinct SNe Ia in the redshift range $0.001 < z < 2.26$. In this sample, we also considered SH0ES Cepheid host distances \cite{sh0es}, which can be used to calibrate the SNe Ia sample. The second dataset are 32 Hubble parameter data compiled by \cite{MorescoEtAl22}, consisting of 32 data points measured with the differential age method, known as cosmic chronometers. This sample also includes a full estimate of non-diagonal terms in the covariance matrix, with systematics contributions.} {The third dataset consists of Baryon Acoustic Oscillation measurements from surveys such as SDSS and DES, which have provided 18 determinations of the angular diameter distance across the redshift range $0.11 < z < 2.4$, as compiled in \cite{Staicova_2022} (Table I). These measurements are expressed in the form $d_A(z)/r_d$, where $r_d$ denotes the sound horizon scale. As argued by \cite{Staicova_2022}, the horizon scale can be considered model-independent, thus they consider $\frac{c}{H_0r_d}$ as a nuisance parameter and marginalize the posterior over it. This model-independent approach turns this BAO dataset suitable for kinematical analysis as we shall perform here.}

The constraints over the free parameters were obtained by sampling the combined likelihood function $\like \propto e^{-\frac{1}{2}\chi ^2}$ through the Affine Invariant method of Monte Carlo Markov Chain (MCMC) analysis, implemented in {\sffamily Python} language by using {\sffamily emcee} software. (See  \cite{GoodmanWeare,ForemanMackey13} for further details). The reconstructions differ from each other by the kind of parametrization. All of them use all data available described earlier.

The $\chi^2$ for some observable $\mathcal{O}$ is described by
\be
\chi^2_{\mathcal{O}}=\mathbf{\Delta \mathcal{O}}^T\cdot\mathbf{C}^{-1}_{\mathcal{O}}\cdot\mathbf{\Delta \mathcal{O}},
\label{errquad}
\ee
where $\mathbf{\Delta \mathcal{O}}= \mathbf{\mathcal{O}}_{\text{model}}-\mathbf{\mathcal{O}}_{\text{data}}$ is the vector distance between the regression model ($\mathbf{\mathcal{O}}_{\text{model}}$) and the data set ($\mathbf{\mathcal{O}}_{\text{data}}$). $\textbf{C}_{\mathcal{O}}$ is a covariance matrix that contains systematic uncertainties of the data set \cite{Betoule,pantheon}. Following the kinematic methodology \cite{JesusEtAl18,JesusEtAl19}, one expects that the cosmological observable $\mathcal{O}$ can be described by a smooth function of the redshift, at least in a low redshift interval. In the present work, $\mathbf{\mathcal{O}}_{\text{model}} = \mathbf{\mathcal{O}}_{\text{model}}(z,\mathbf{p})$ will be a polynomial in $z$ with free parameters encapsulated in $\mathbf{p}$.

\subsection{Comoving Distance parameterization}

In order to reconstruct the normalized torsion function (\ref{Phisolf}) by using Supernovae and $H(z)$ data, we use the relation of Comoving Distance $D_C$ with magnitude $m$ {\cite{WangY2009,Condon:2018eqx}}:
\be
m = 5\log_{10}\left[ (1+z) D_C \right]+\mathcal{M},
\label{mmm}
\ee
where $\mathcal{M}$ is a constant, and write the parameterization of $D_C$ as:
\be
D_C = z+d_2z^2+d_3z^3\,,
\label{pold}
\ee
with $d_2$ and $d_3$ free parameters to be constrained, together with the Hubble parameter $H_0$. {Using SNe Ia data for the magnitudes \cite{pantheon+} and Eqs. \eqref{errquad}-\eqref{pold}, $\chi^2_{m}$ is calculated. In the case of the dataset that we have chosen to work with, Pantheon+\&SH0ES, we have one additional parameter, the SNe Ia absolute magnitude, $M$, which is included in constant $\mathcal{M}$ above. Thus we have $\mathcal{O} = m$ and $\mathbf{p} = (M,H_0,\,d_2,\,d_3)$.}

By using \eqref{conectaHD} with Eq. \eqref{pold} to obtain $E(z)$, we have:
\be
E(z) = \frac{1}{1+2d_2 z+3d_3 z^2}\,,
\label{pold2}
\ee
so that $\mathcal{O} = H$ with $\mathbf{p} = (H_0,\,d_2,\,d_3)$. By using the cosmic chronometers $H(z)$ data from \cite{Magana2018},  Eq. \eqref{errquad} can be combined with Eq. \eqref{pold2}, and $\chi^2_{H}$ is calculated.

{The angular diameter distance to be constrained by BAO data relates with comoving distance by:
\be
d_A(z)=\frac{c}{H_0}\frac{D_C(z)}{1+z}
\label{dADC}
\ee
}

\begin{figure}[h!]
    \centering
    \includegraphics[width=.9\textwidth]{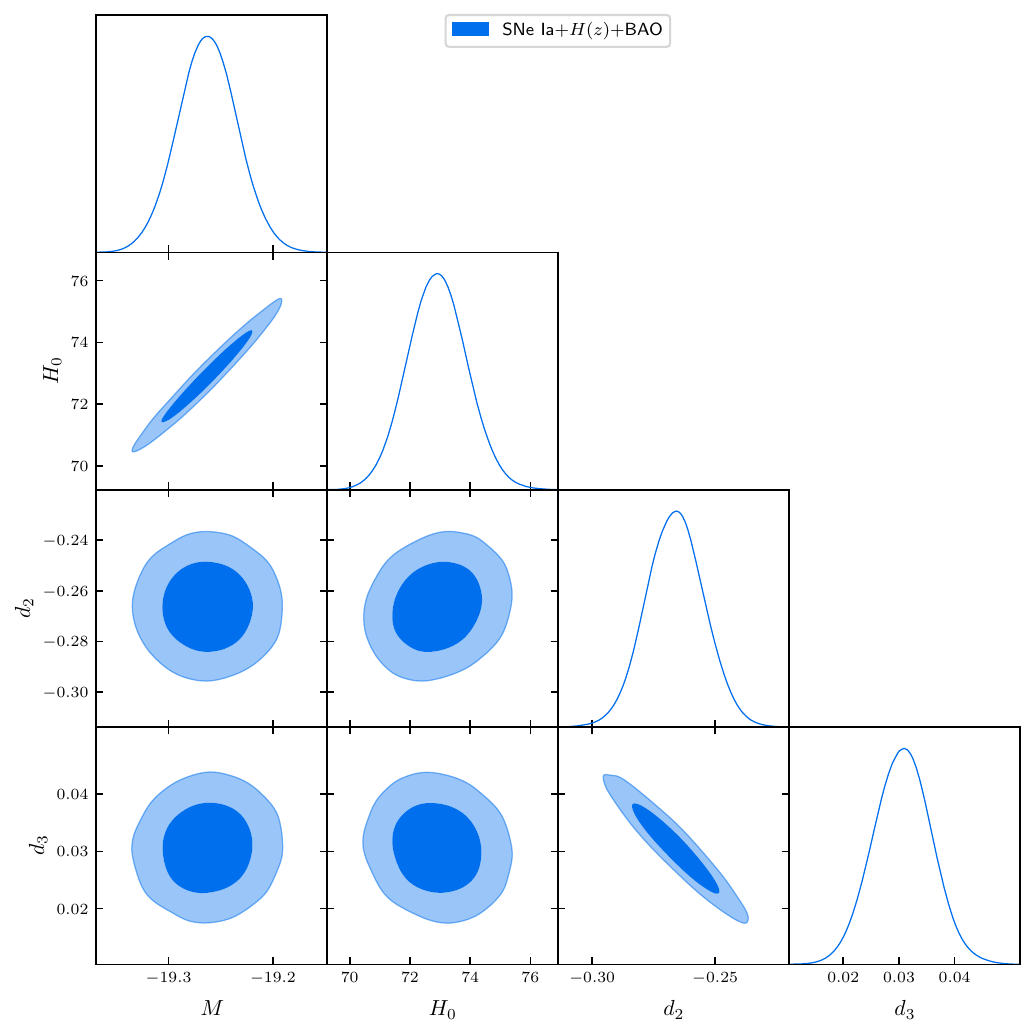}
    \caption{Contours for the joint analysis of SNe Ia, $H(z)$ {and BAO} data at 1$\sigma$ and 2$\sigma$ for the free parameters in the Comoving Distance parameterization (\ref{pold}).}
    \label{PCPP1}
\end{figure}

{From this, one can build the $\chi^2_\text{BAO}$ \cite{Staicova_2022} and by probing the likelihood, $\like \propto e^{-\frac{1}{2}(\chi^2_{H}+\chi^2_{m}+\chi^2_\text{BAO})}$, the confidence contours} for the free parameters $M$, $H_0$, $d_2$ and $d_3$ were generated and are shown in Figure \ref{PCPP1}.
\newpage

{As can be seen on this Figure,} the parameters are well constrained for the SNe Ia+$H(z)$ combined dataset. One can also realize that there is some correlation between parameters $d_2$ and $d_3$ {and between $H_0$ and $M$, as expected, while there is almost no correlation among the other parameters.} One can also realize that none of the parameters are negligible, as they are not compatible with zero, at least at 95\% c.l.

\begin{figure}[ht]
    \centering
    \includegraphics[width=.8\textwidth]{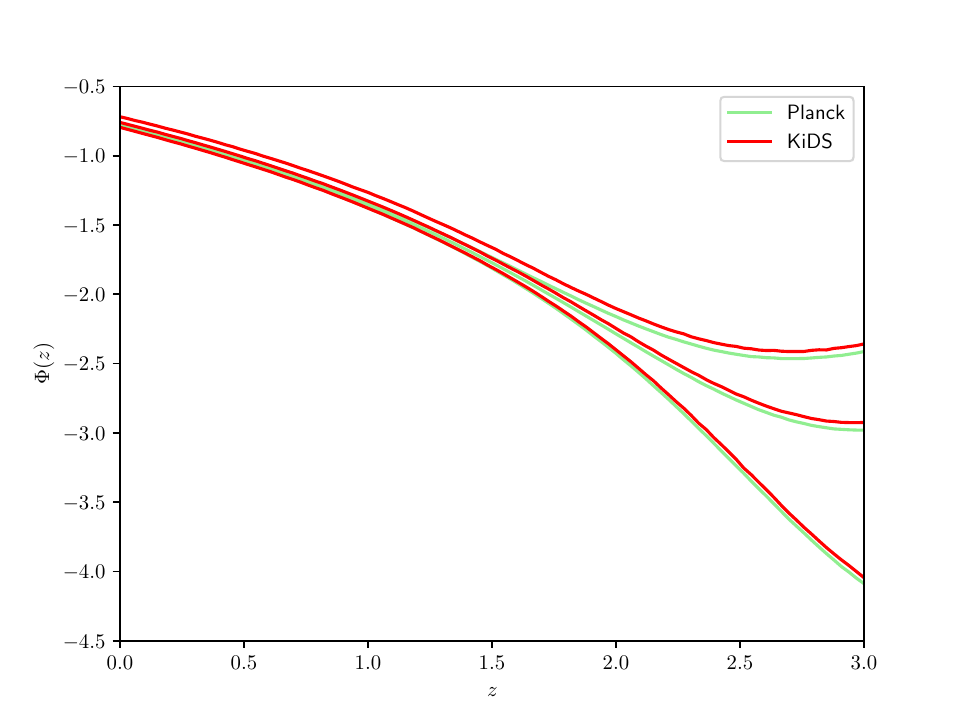}
    \caption{{Reconstruction of the normalized torsion function $\Phi$ with the parameterization (\ref{pold}) for the Comoving Distance, with a 3$\sigma$ Planck prior (green) and with an 1$\sigma$ KiDS prior (red) over $\Omega_m$.}}
    \label{PCPP2}
\end{figure}

From the MCMC chains corresponding to the parameters $(H_0,d_2,d_3)$ and with the relations \eqref{pold} and \eqref{pold2}, we wanted to reconstruct the normalized torsion function $\Phi$ from the relation \eqref{Phisolf}. {However, as can be seen in \eqref{Phisolf}, another parameter is needed to obtain the reconstruction of $\Phi(z)$, namely, $\Omega_{m0}$. Since it is a dynamic parameter that can not be obtained from kinematic parametrizations, we chose to work with a Gaussian prior, $\Omega_m=0.315\pm0.021$, which corresponds to 3$\sigma$ from the Planck estimate \cite{Planck2018}. Furthermore, aiming for model-independence, as the Planck result relies on the $\Lambda$CDM model, we also work with a prior over $\Omega_m$ from cosmic shear measurements, namely, the KiDS-1000 survey \cite{Kids}, of $\Omega_m=0.270\pm0.079$\footnote{Actually, for simplicity, this is a symmetrization of the KiDS result, $\Omega_m=0.270^{+0.056}_{-0.102}$, following the approach suggest by D'Agostini \cite{DAgostini}.}.} The result of this reconstruction is shown in Figure \ref{PCPP2}.

As can be seen from this Figure, in the context of this torsion model, the torsion can not be neglected at the considered redshift interval, which corresponds to the $H(z)$ and SNe Ia data redshifts.

The mean values of the free parameters are summarized in the table below:

\begin{table}[ht]
    \centering
    \begin{tabular} { c | c}

 Parameter &  68\% and 95\% limits\\
\hline
{\boldmath$M              $} & $-19.262^{+0.028+0.056}_{-0.028-0.058} $\\

{\boldmath$H_0            $} & $72.92\pm0.96\pm1.9        $\\

{\boldmath$d_2            $} & $-0.267\pm0.012\pm0.024  $\\

{\boldmath$d_3            $} & $0.0308\pm0.0053\pm0.011 $\\
\hline
\hline
\end{tabular}
    \caption{Mean values with 68\% and 95\% c.l. {constraints for the parameters $M$,} $H_0$, $d_2$ and $d_3$ for the Comoving Distance parameterization.}
    \label{tab1}
\end{table}

We can see from this Table that these results are compatible with other results in the literature, as for instance, \cite{JesusEtAl18}, where it has been used the same parametrization, Eq. \eqref{pold}, and obtained $H_0=69.1\pm1.5$ km/s/Mpc, $d_2=-0.253\pm0.016$ and $d_3=0.0299\pm0.0044$, at 1$\sigma$ c.l., which are all compatible within $1\sigma$ c.l. {This result is compatible with the local result from SH0ES \cite{SH0ES22}, $H_0=73.04\pm1.04$ km/s/Mpc, as expected, since we are taking into account the Pantheon+\&SH0ES data.}


\subsection{\texorpdfstring{$H(z)$}{H(z)} parametrization}

Alternatively, we can also reconstruct $\Phi$ by means of the parameterization of the Hubble parameter:
\be
H(z) = H_0(1+h_1z+h_2z^2)\,,
\label{pold4}
\ee
with $\mathcal{O} = H$ and $\mathbf{p} = (H_0,\,h_1,\,h_2)$.
By using the $H(z)$ data from cosmic chronometers \cite{MorescoEtAl22}, Eq. \eqref{errquad} combined with Eq. \eqref{pold4}, the $\chi^2_{H}$ function can be calculated. By using \eqref{conectaHD} with Eq. \eqref{pold4}, the comoving distance is:
\be
D_C(z)\equiv\int_0^z\frac{dz}{1+h_1 z+h_2 z^2},
\label{DcHz}
\ee
which can be combined with Eq. \eqref{mmm} to calculate $\chi^2_m$ through SNe Ia data. {From this, $d_A$ can be obtained by \eqref{dADC} and $\chi^2_\text{BAO}$ can be calculated.}

\begin{figure}[ht]
    \centering
    \includegraphics[width=.9\textwidth]{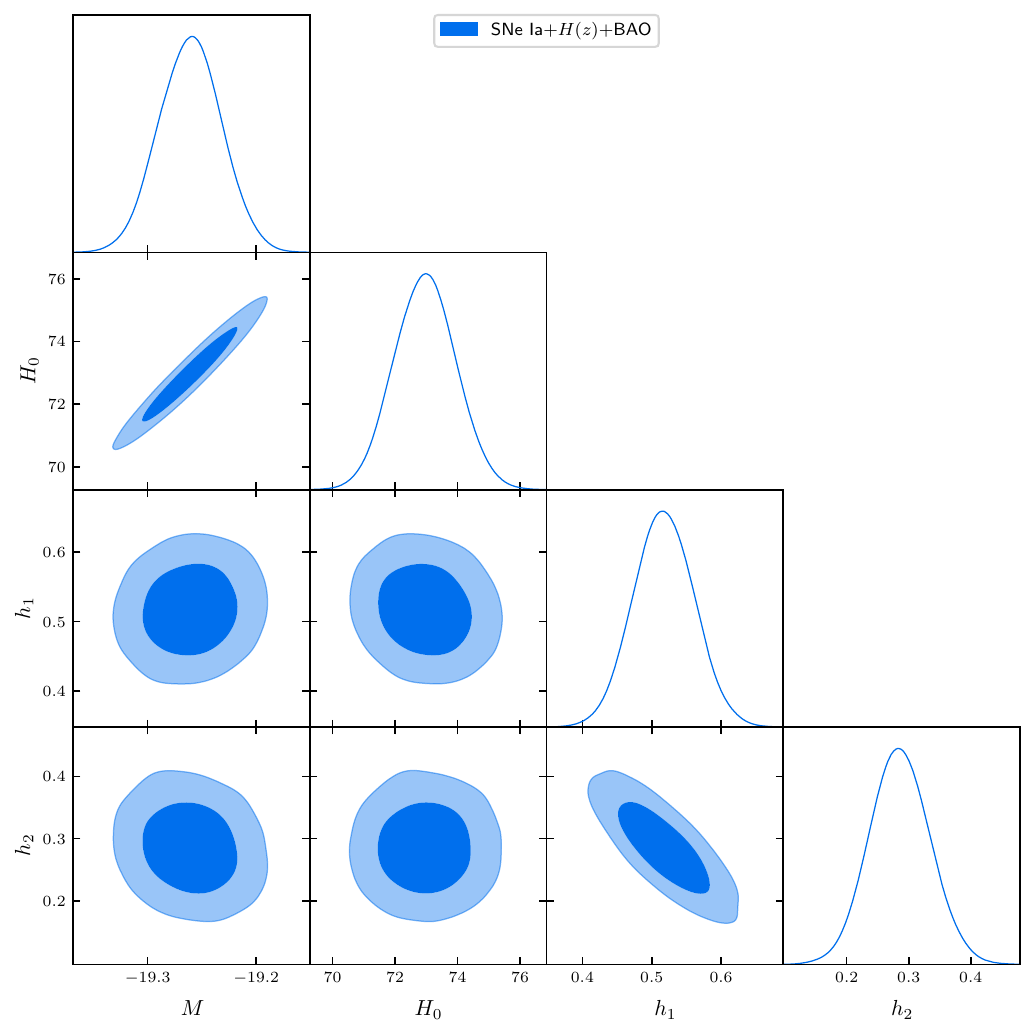}
    \caption{Contours for the joint analysis of SNe Ia,  $H(z)$ {and BAO} data at 1$\sigma$ and 2$\sigma$ for the free parameters in the Hubble parameter parameterization (\ref{pold4}).}
    \label{PCPP3}
\end{figure}

{The likelihood  function $\like \propto e^{-\frac{1}{2}(\chi^2_{H}+\chi^2_{m}+\chi^2_\text{BAO})}$ furnishes} the contours of the parameters $H_0$, $h_1$ and $h_2$ for 1$\sigma$ and $2\sigma$ c.l., as shown in Figure \ref{PCPP3}.

As can be seen from this Figure, the free parameters are well constrained by the combination of SNe Ia+$H(z)${+BAO} data. The main correlation is between parameters $h_1$ and $h_2$. There is also some correlation between $H_0$ and $h_2$. The mean values and 68\% and 95\% c.l. for the free parameters are presented in Table \ref{tab2}.

\begin{table}[ht]
    \centering
    \begin{tabular} { c | c}

 Parameter &  68\% and 95\% limits\\
\hline
{\boldmath$M             $} & $-19.260\pm0.028\pm0.057$\\

{\boldmath$H_0            $} & $72.98^{+0.97+2.0}_{-0.97-1.9}$\\

{\boldmath$h_1            $} & $0.517^{+0.044+0.090}_{-0.044-0.088}$\\

{\boldmath$h_2            $} & $0.286^{+0.050+0.10}_{-0.050-0.098   }$\\
\hline
\hline
\end{tabular}
    \caption{Mean values and 68\% and 95\% c.l. {constraints for the parameters $M$,} $H_0$, $h_1$ and $h_2$ for the Hubble parameter parameterization.}
    \label{tab2}
\end{table}

\newpage

The results shown in Table \ref{tab2} can be compared with a previous analysis in the context of the same $H(z)$ parametrization. In \cite{JesusEtAl18}, they have obtained $H_0=68.8\pm1.6$ km/s/Mpc, $h_1=0.522\pm0.065$ and $h_2=0.192\pm0.026$. $H_0$ and $h_1$ are compatible within 1$\sigma$ c.l., while $h_2$ is compatible only at 2$\sigma$ c.l. {This result for $H_0$ is also compatible with the local result from SH0ES \cite{SH0ES22} alone, $H_0=73.04\pm1.04$ km/s/Mpc, within $2\sigma$ c.l., as expected, since our analysis involves the Pantheon+\&SH0ES data.}

\begin{figure}[ht]
    \centering
    \includegraphics[width=.8\textwidth]{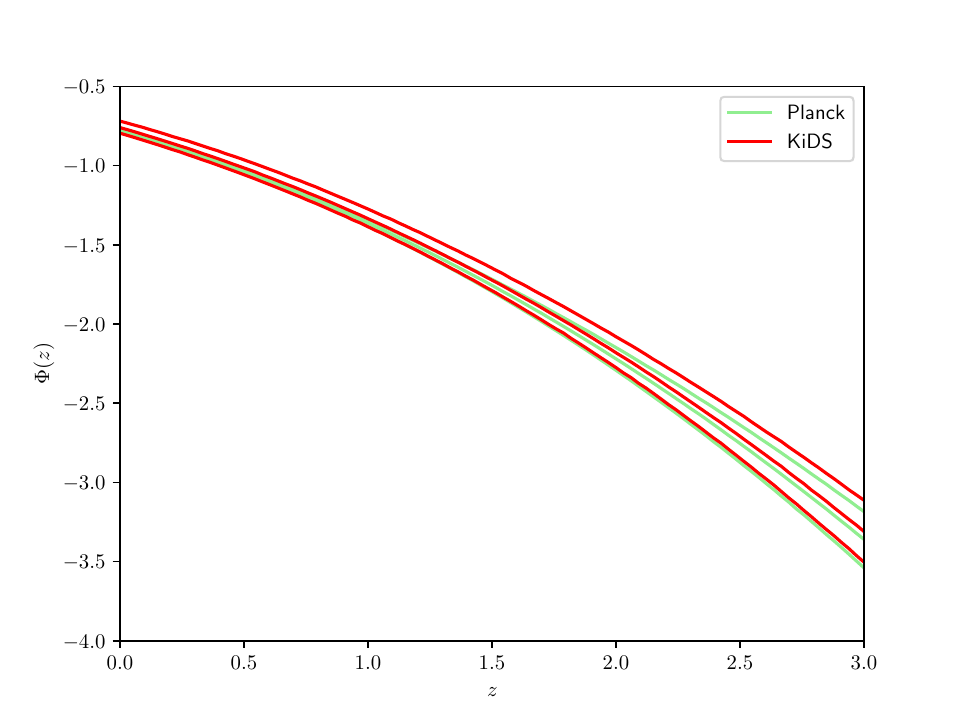}
    \caption{{Reconstruction of the normalized torsion function $\Phi$ with the parameterization (\ref{pold4}) for the Hubble parameter $H(z)$, with a 3$\sigma$ Planck prior (green) and with an 1$\sigma$ KiDS prior (red) over $\Omega_m$.}}
    \label{PCPP}
\end{figure}

From the MC chains corresponding to the parameters $(H_0,h_1,h_2)$ and with the relations \eqref{DcHz} and \eqref{pold4}, we reconstructed the normalized torsion function $\Phi$ from the relation \eqref{Phisolf}. We have also used the 3$\sigma$ Planck prior, $\Omega_{m0}=0.315\pm0.021$. This is shown in Figure \ref{PCPP}.

\newpage

As can be seen from this Figure, the torsion is well reconstructed from the comoving distance parametrization.

\begin{figure}[ht]
    \centering
    \includegraphics[width=.49\textwidth]{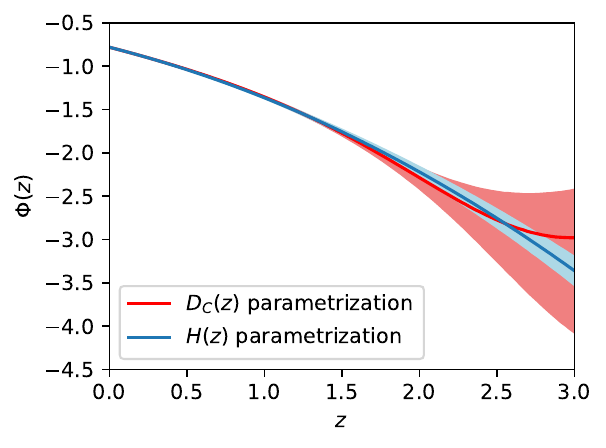}
    \includegraphics[width=.49\textwidth]{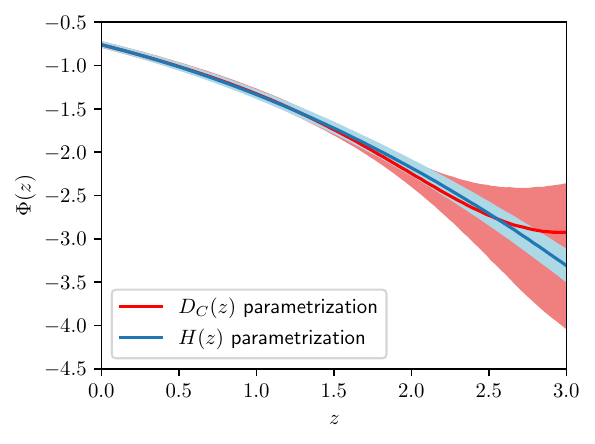}
    \caption{Superposition of the two different reconstructions for $\Phi$ within $1\sigma$ c.l. {\textbf{Left:} 3$\sigma$ Planck prior over $\Omega_m$. \textbf{Right:} 1$\sigma$ KiDS prior over $\Omega_m$.}}
    \label{PCPP45}
\end{figure}

The concordance between the two reconstructions above can be seen in Figure \ref{PCPP45}, where the $1\sigma$ interval for both cases are shown in the same figure. We see that both reconstructions are in good agreement within $1\sigma$ c.l.

\newpage

\begin{figure}[ht]
    \centering
    \includegraphics[width=0.49\linewidth]{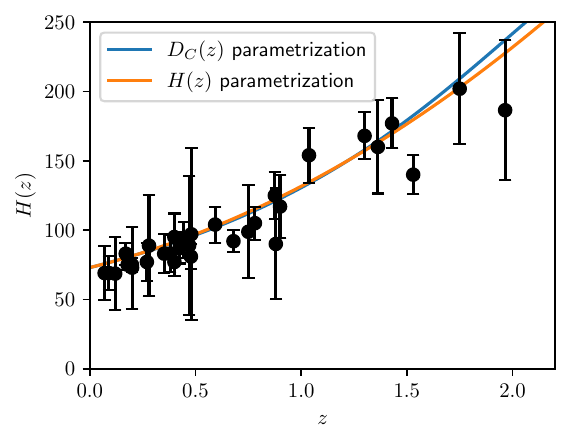}
    \includegraphics[width=0.49\linewidth]{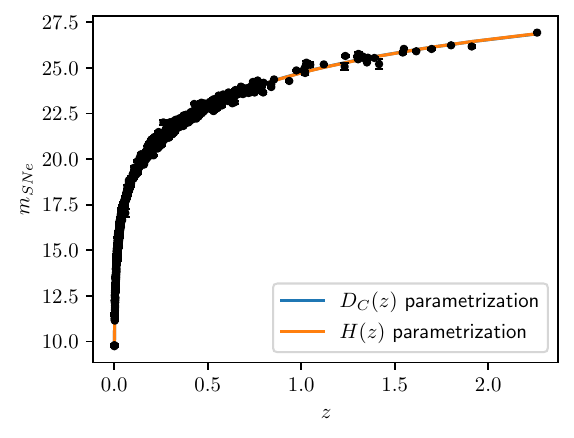}
    \includegraphics[width=0.49\linewidth]{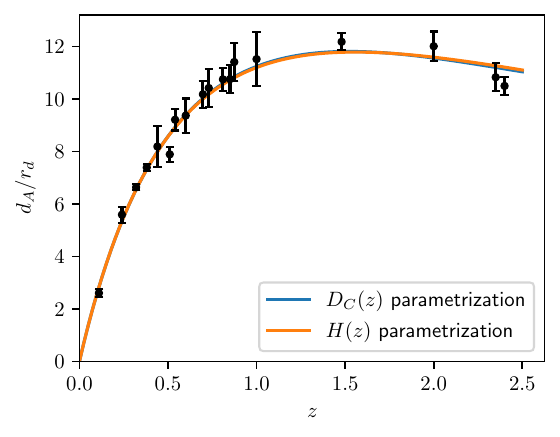}
    \caption{\textbf{Upper Left:} $H(z)$ data are shown together with best-fit $D_C(z)$ and $H(z)$ parametrizations. \textbf{Upper Right:} Pantheon+\&SH0ES SNe Ia magnitudes are shown together with best-fit $D_C(z)$ and $H(z)$ parametrizations. \textbf{Bottom:} {$d_A(z)/r_d$ data are shown together with best-fit $D_C(z)$ and $H(z)$ parametrizations.}}
    \label{dataplots}
\end{figure}

{In Fig. \ref{dataplots}, we show $H(z)$, Pantheon+\&SH0ES and BAO data together with the best fit models as given by Tables \ref{tab1} and \ref{tab2}.} As can be seen from these Figures, both parametrizations present good fits to data in the data redshift interval. {As one may see in the upper left panel, $H(z)$ data have large error bars and low scattering around the best-fit curves. It may be related to a possible overestimation of $H(z)$ uncertainties, which has already been pointed out by \cite{JesusEtAl18HzCor,KvintEtAl25}.}


\subsection{Deceleration Parameter}
In order to validate the discussed torsion model as a candidate to describe the dark energy sector, we reconstruct the deceleration parameter $q$ in order to compare to the standard model one. From (\ref{qa}) we obtain:
\be
q(z) = -1+\left( 1+z \right) \dfrac{H^\prime}{H}\,.
\label{pold6}
\ee

Using this expression, it is easy to show that for the comoving distance parametrization \eqref{pold}, with $E(z)$ given by Eq. \eqref{pold2}, the deceleration parameter is given by:
\be
q(z)=-1-\frac{(1+z)(2d_2+6d_3z)}{1+2d_2z+3d_3z^2}
\ee

Similarly, it is easy to show that for the $H(z)$ parametrization \eqref{pold4}, the deceleration parameter is given by:
\be
q(z)=-1+\frac{(1+z)(h_1+2h_2z)}{1+h_1z+h_2z^2}
\ee

\begin{figure}[ht]
    \centering
    \includegraphics[width=.8\textwidth]{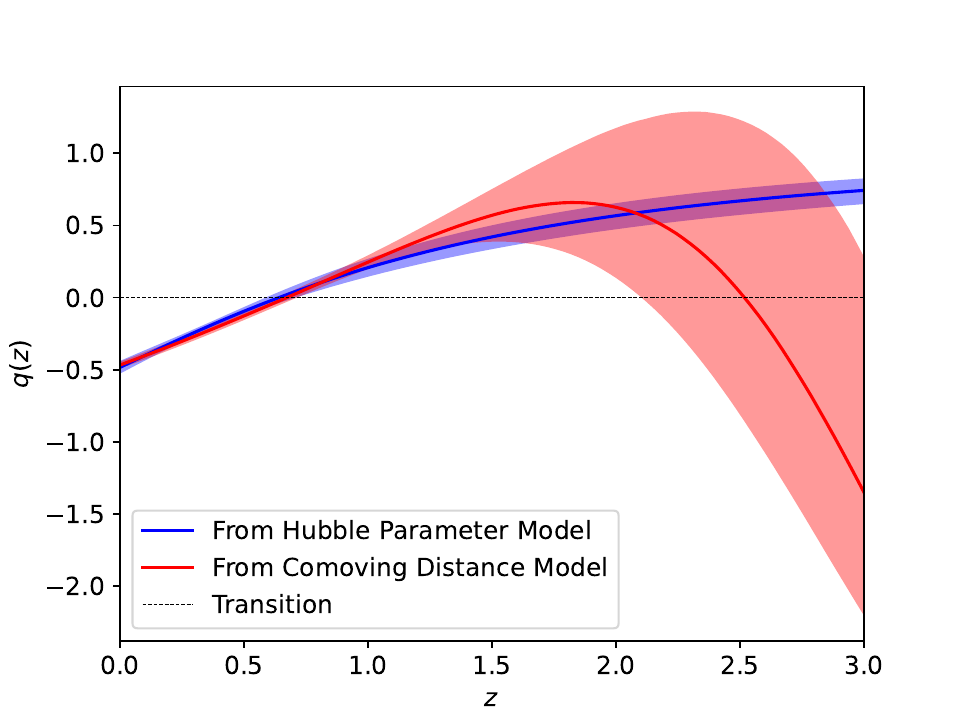}
    \caption{{Reconstruction of the deceleration parameter $q(z)$ for the two parameterizations:  $D_C(z)$ and $H(z)$, both {with 1$\sigma$ confidence interval.}}}
    \label{lowred}
\end{figure}

Having the MCMC chains of the parameters $d_2,\,d_3$ and $h_1,\, h_2$ for both parametrizations, these expressions can be used to reconstruct $q(z)$ and the results are plotted, showing 1$\sigma$ confidence intervals, in Figure \ref{lowred}.

It is evident from this figure that the transition redshift $z_t$ occurs at about $0.60 - 0.75$, in good agreement, for instance, with the value of $z_t \simeq 0.67$ for the standard $\Lambda$CDM model with $\Omega_m \simeq 0.3$ and $\Omega_\Lambda \simeq 0.7$.

\subsection{Comparison with \texorpdfstring{$\Lambda$}{L}CDM}
In order to test the viability of the present models according to the current observations, here we compare them with the standard cosmological model, flat $\Lambda$CDM. {For this aim, we use the so called Bayesian Information Criterion, BIC, which is an approximation of the Bayesian Evidence and penalizes the excess of parameters \cite{Schwarz78,Liddle04}. The number of data is $n_{data}=n_{SN}+n_{CC}+n_{BAO}=1707$ for all models.}

\begin{table}[h!]
\begin{tabular}{|l|c|c|c|c|c|}
\hline
Model            & $\chi^2_{min}$ & $n_{par}$ & $\chi^2_\nu$ & BIC      & $\Delta$BIC\\ \hline
Quadratic $H(z)$ & 1492.115       & 3         & 0.8762       & 1521.885 & +6.636\\ \hline
Cubic $D_C(z)$   & 1492.296       & 3         & 0.8763       & 1522.066 & +6.817\\ \hline
$\Lambda$CDM     & 1492.921       & 2         & 0.8761       & 1515.249 & 0\\ \hline
\end{tabular}
\caption{Comparison among the 2 cosmographic models studied here and the standard cosmological model.}
    \label{tabBIC}
\end{table}

{As one may see in this table, while the cosmographic models have lower $\chi^2_{min}$ and {similar} $\chi^2_\nu$, they are strongly penalized for having 1 parameter more than $\Lambda$CDM, so that they present a higher BIC.}

\section{Concluding remarks}

We have studied the torsion effects in Cosmology as a candidate to dark energy in the Universe. The analysis was done by assuming that the torsion function $\phi(t)$ drives the recent cosmic evolution, and contrary to recent works where the specific form of the torsion function were fixed, here we have reconstructed the torsion function by using observational data of Supernovae and Hubble parameter measurements. The method of reconstruction was based on a cosmographic analysis, where previous parametrizations for the comoving distance and the Hubble parameter are adopted and the free parameters of the models are constrained by observational data. The values of the free parameters are then used to reconstruct the normalized torsion function $\Phi(z)$, without to appeal to a specific cosmological model. The value of the $H_0$ parameter (within $2\sigma$ c.l.) obtained by the two parametrizations are in good agreement to both recent estimates, namely measures obtained by the local distance ladder and by the cosmic microwave background power spectrum. However, there is a slight preference for the later one, as it is compatible within 1$\sigma$, while the first one is compatible within 2$\sigma$ only. The estimate of the deceleration parameter is also in good agreement to the standard model estimate. Such interesting features show that torsion effect in the evolution of the universe must be further investigated.

\appendix
\section{{The Einstein-Cartan equations}}

Here we briefly present the construction of the simplest model of gravity with torsion, the so called Einstein-Cartan (EC) gravity. We follow the same notation of \cite{Kranas:2018jdc} and motivation of \cite{Shapiro:2001rz}. 

We start with the construction of the covariant derivative in General Relativity.
It is well known that the action of the covariant derivative in a four vector $A^\alpha$ is given by:
\begin{equation}
    \tilde{\nabla}_\beta A^\alpha = \partial_\beta A^\alpha + \tilde{\Gamma}^\alpha_{~~\beta\gamma}A^\gamma \,,
\end{equation}
where $\Tilde{\Gamma}^\alpha_{~~\beta\gamma}$ defines a connection which transforms in a special non-tensor way in order to assure the covariant derivative to be a tensor. However, such connection contains some ambiguity since that $\Tilde{\Gamma}^\alpha_{~~\beta\gamma}$ remains a
tensor if one adds to it any tensor $C^\alpha_{~~\beta\gamma}$, namely $\Tilde{\Gamma}^\alpha_{~~\beta\gamma}\to \Tilde{\Gamma}^\alpha_{~~\beta\gamma} + C^\alpha_{~~\beta\gamma}$. In General Relativity a very special choice of the connection is done, by imposing the symmetry on the lower indices, $\Tilde{\Gamma}^\alpha_{~~\beta\gamma}=\Tilde{\Gamma}^\alpha_{~~\gamma\beta}$, and the so called metricity condition, $\tilde{\nabla}_\alpha g_{\mu\nu}=0$. If these two conditions are satisfied, the unique solution for $\Tilde{\Gamma}^\alpha_{~~\beta\gamma}$ is:
\begin{equation}
    \Tilde{\Gamma}^\alpha_{~~\beta\gamma}=\frac{1}{2}g^{\alpha\lambda}(\partial_\beta g_{\lambda\gamma}+\partial_\gamma g_{\lambda\beta}- \partial_\lambda g_{\beta\gamma})\,,\label{A2}
\end{equation}
which is called Christoffel symbol, a particular case of the affine connection that depends on the metric only. Indeed, it is the simplest affine connection among several others.

The simplest extension from the standard case (\ref{A2}) can be achieved by assuming the existence of a new connection ${\Gamma}^\alpha_{~~\beta\gamma}$ that is non-symmetric:
\begin{equation}
    {\Gamma}^\alpha_{~~\beta\gamma}-{\Gamma}^\alpha_{~~\gamma\beta}\equiv 2{S}^\alpha_{~~\gamma\beta}\,,\label{A3}
\end{equation}
but still maintains the metricity condition, ${\nabla}_\alpha g_{\mu\nu}=0$. In this case the tensor ${S}^\alpha_{~~\gamma\beta}$ is known as torsion. Notice that we are using (\ref{A2}) for the Christoffel symbol and the notation without a tilde for the connection and for the corresponding covariant derivative with torsion.

In order to satisfy (\ref{A3}) and the metricity condition, the connection can be written as:
\begin{equation}
    \Gamma^\alpha_{~~\mu\nu}=\Tilde{\Gamma}^\alpha_{~~\mu\nu}+K^\alpha_{~~\mu\nu}\,,\label{A4}
\end{equation}
where  $K^\alpha_{~~\mu\nu}$ defines the contorsion tensor, which can be written in terms of the torsion tensor $S^\alpha_{~~\mu\nu}$ as:
\begin{equation}
    K^{\alpha}_{~~\mu\nu} = S^{\alpha}_{~~\mu\nu} + S_{\mu\nu}^{~~~\alpha} + S_{\nu\mu}^{~~~\alpha} \,.\label{K}
\end{equation}
With (\ref{A4}) and (\ref{K}), all the objects of the standard Riemannian geometry can be constructed, as the Ricci tensor:
\begin{equation}
    R_{\mu\nu} = -\partial_\nu\Gamma^\alpha_{~\mu\alpha} + \partial_\alpha\Gamma^\alpha_{~\mu\nu} - \Gamma^\beta_{~\mu\alpha}\Gamma^\alpha_{~\beta\nu} + \Gamma^\beta_{~\mu\nu}\Gamma^\alpha_{~\beta\alpha}\,,
\end{equation}
and the Ricci scalar, $R=R_{\mu\nu}g^{\mu\nu}$. Thus, all the equations of gravitation in EC framework maintain the same form as the standard one in General Relativity. {The Einstein-Hilbert action is generalised to the EC gravity as \cite{Ivanov2016,Shapiro:2001rz}:
\begin{equation}
    S = -\frac{1}{\kappa^2}\int d^4x \sqrt{-g}(R-2\Lambda)\,,
\end{equation}
with $\kappa^2 = {8\pi G}$ and $\Lambda$ the cosmological constant term.
} The Einstein equation is:
\begin{equation}
    R_{\mu\nu} - \frac{1}{2}R g_{\mu\nu} + \Lambda g_{\mu\nu}=\kappa^2 T_{\mu\nu} \,,\label{Rmunu}
\end{equation}
with $T_{\mu\nu}$ the energy momentum tensor. 

Finally, in order to maintain the homogeneity and isotropy of the 3-space in a FRW background, it was showed by \cite{Tsamparlis:1981xm} that the torsion tensor must be proportional to a scalar function of time only, $\phi(t)$, with the non-null components given by:

\begin{equation}
    S^1_{~01} = S^2_{~02} = S^3_{~03} = \phi,
\end{equation}
and
\begin{equation}
    S^1_{~10} = S^2_{~20} = S^3_{~30} = -\phi.
\end{equation}
In a FRW metric, $\mathrm{d}s^{2}= -\mathrm{d}t^{2}+ a^{2}\left[(1-Kr^{2})^{-1}\mathrm{d}r^{2} +r^{2}\mathrm{d}\theta^{2} +r^{2}\sin^{2}\theta\mathrm{d}\varphi\right]$, after a lengthy calculation, the components of the connection (\ref{A4}) are given by:
\begin{align}
& \Gamma^{0}{}_{11}= \frac{a\dot{a}}{1-Kr^{2}}+\frac{2\phi a^{2}}{1-Kr^{2}}\,,\quad
\Gamma^{0}{}_{22}= r^{2}a\dot{a}+2\phi r^{2} a^{2}\,,\quad
\Gamma^{0}{}_{33}=  a\dot{a}r^{2}\sin^{2}\theta+2\phi
a^{2}r^{2}\sin^{2}\theta\,,  \notag \\ & \Gamma^{1}{}_{01}= \frac{\dot{a}}{a}+ 2\phi\,,\quad \Gamma^{1}{}_{10}= \frac{\dot{a}}{a}\,,\quad \Gamma^{1}{}_{11}= \frac{Kr}{1-Kr^{2}}\,,  \notag\\ & \Gamma^{1}{}_{22}= -r\left(1-Kr^{2}\right)\,,\quad \Gamma^{1}{}_{33}= -r\left(1-Kr^{2}\right)\sin^{2}\theta\,,  \notag\\ & \Gamma^{2}{}_{02}= \frac{\dot{a}}{a}+ 2\phi\,,\quad \Gamma^{2}{}_{20}= \frac{\dot{a}}{a}\,,\quad \Gamma^{2}{}_{12}= \Gamma^{2}{}_{21}= \frac{1}{r}\,,\quad \Gamma^{2}{}_{33}= -\cos\theta\sin\theta\,, \notag\\ & \Gamma^{3}{}_{03}= \frac{\dot{a}}{a}+ 2\phi\,,\quad \Gamma^{3}{}_{30}= \frac{\dot{a}}{a}\,,\quad \Gamma^{3}{}_{13}= \Gamma^{3}{}_{31}= \frac{1}{r}\,,\quad \Gamma^{3}{}_{23}= \Gamma^{3}{}_{32}= \cot\theta\,,  \label{Gammas}
\end{align}
from which is evident that the second terms appearing in $\Gamma^0_{~ii}$ and $\Gamma^i_{~0i}$ proportional to $\phi$ comes from the contorsion tensor in (\ref{A4}) while the first terms are the usual Christoffel symbols. Finally, the Friedmann equations that follows from (\ref{Rmunu}) are given by (\ref{H2})-(\ref{Hd2}), with $K=0$. 

\begin{acknowledgments}
The authors sincerely thank the referee for their insightful comments and suggestions, which have significantly improved the quality of this work. This study was financed in part by the Coordena\c{c}\~ao de Aperfei\c{c}oamento de Pessoal de N\'ivel Superior - Brasil (CAPES) - Finance Code 001. JFJ acknowledges financial support from  {Conselho Nacional de Desenvolvimento Cient\'ifico e Tecnol\'ogico} (CNPq)  (No. 314028/2023-4). SHP acknowledges financial support from  CNPq  (No. 308469/2021).
\end{acknowledgments}


\end{document}